# Experiments with biased side electrodes in electron cyclotron resonance ion sources[a)]

A. G. Drentje,[1,b)] A. Kitagawa,[1] T. Uchida,[2] R. Rácz,[3] and S. Biri[3]

[1]*National Institute of Radiological Sciences (NIRS), 4-9-1 Anagawa, Inage-ku, Chiba-shi 263-8555, Japan*

[2]*Bio-Nano Electronics Research Centre, Toyo University, 2100 Kujirai, Kawagoe-shi 350-8585, Japan*

[3]*Institute for Nuclear Research (Atomki), H-2026 Debrecen, Bem ter 18/c, Hungary*

The output of highly charged ions from an electron cyclotron resonance ion source (ECRIS) consists of ionic losses from a highly confined plasma. Therefore, an increase of the output of the ions of interest always is a compromise between an increase in the confinement and an increase of the losses. One route towards a solution consists of attacking the losses in directions – i.e., radial directions – that do not contribute to the required output. This was demonstrated in an experiment (using the Kei ECRIS at NIRS, Japan) where radial losses were electrostatically reduced by positively biasing one set of six "side" electrodes surrounding the plasma in side-ward directions attached (insulated) to the cylindrical wall of the plasma chamber. Recently new studies were performed in two laboratories using two essentially different ion sources. At the BioNano ECRIS (Toyo University, Japan) various sets of electrodes were used; each of the electrodes could be biased individually. At the Atomki ECRIS (Hungary), one movable, off-axis side electrode was applied in technically two versions. The measurements show indeed a decrease of ionic losses but different effectivities as compared to the biased disk.

## I. INTRODUCTION

The plasma in electron cyclotron resonance ion sources (ECRISs) is generally confined inside a magnetic configuration that forms a so-called minimum B ($B_{min}$)-configuration with as primary task to (magnetically) confine electrons. The total charge of the electrons must be compensated by the total charge of the ions; thus the ions are being confined electrostatically. In order to create a high density of highly charged ions (HCI), the ion confinement should be long enough, such that by sequential electron impact ionization the high charge states can be formed. The ions can be extracted because the $B_{min}$-configuration has some leakage (*non-perfect confinement*). Users are interested only in those ions that are leaking out at the axis of the source at the extraction side. Source engineers are spending many efforts[1] to find a best compromise between optimizing these leakage ion-currents and at the same time reducing the leakage for best confinement.

One "trivial" solution is a reduction of the leakage currents in the other possible directions: radially in the six directions of the poles of the (radial) hexapole magnet and axially in the so-called injection direction. Our base statement is that "axial" particle losses are dominated by fluxes of electrons while "radial" losses are dominated by fluxes of ions because of easier electron transport (as compared to ion transport) along the magnetic field lines and the opposite situation perpendicular to it. Experimentally, these facts have been demonstrated already in Grenoble.[2] The same laboratory also started with the application of a biased electrode on axis at the injection side, nowadays known as "biased disk" (BD). The BD showed to be very effective.[3,4] Various explanations have been made and reviewed.[5] Measurements of radial currents (to an insulated cylinder) and axial currents (to end electrodes) in an ECRIS have been performed also by Stiebing *et al.*[6] in Frankfurt. It should be noted however, that the effect of an axially positioned BD on the plasma seems more complicated: it can decrease electron losses, at the same time, may behave as cavity tuner.[7]

It is interesting to determine in experiments if such a reduction of the ionic losses from the radial leakage areas can be realized. For that, we have performed measurements in three essentially different ECRISs. Earlier we have checked this possibility at NIRS in a small prototype permanent magnet ECRIS intended for production of high intensity beams of carbon ions ($C^{4+}$) to be injected in an accelerator for carbon therapy. A set of six "side" electrodes (individually connected) surrounding the plasma in side-ward directions was attached (insulated) to the cylindrical wall of the plasma chamber at positions corresponding to the poles of the hexapole magnet. Application of a positive voltage of 20 V to the six electrodes resulted in a substantial increase of the highly charged ions output[8] as well as a shift in the charge state distribution (CSD) towards higher charges. Most important observation is that improvement due to the application of the biased cylinder is adding to that of the biased disk.

## II. EXPERIMENTS AT TOYO UNIVERSITY

Application of the technique described above has so far not been made in other ECRISs, perhaps due to lack of space inside a regular plasma chamber and resulting instrumental difficulties. Nevertheless it is interesting to perform experiments. The "BioNano-ECRIS" at Toyo University offered such a possibility. The primary task of this instrument is the usage for a wide range of experiments with all kinds of lowly charged ions including, e.g., fullerene-ions. The configuration is simple, with emphasis on flexibility. For that reason the plasma chamber has a large inside diameter (140 mm). Production of high intensity beams of highly charged ions does not have a high priority.[9]

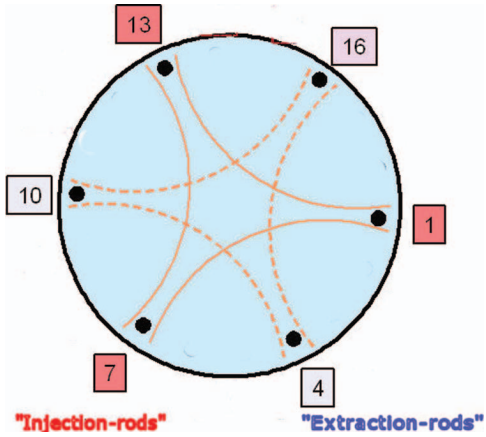

FIG. 1. Cross section of the plasma chamber of Bio-Nano ECRIS showing the rods that can be biased separately.

### A. Six biased rods

Inside this source provisions were made to insert various electrodes, each having an insulated feed-through. In particular along the side wall of the chamber, at a distance of 10 mm, long rods were installed (see Figure 1) at positions corresponding to the location of the poles of the hexapole magnet. At the applied RF frequency of 10 GHz, the ECR resonance zone in the midplane of the magnets is located about 50 mm from the axis; therefore, the rods are about 10 mm from the ECR-surface. This set-up was also planned for a study of carbon contamination[10] of the plasma chamber. In the present discussion, we restrict to measurements where the six side rods and the biased disk were set at various bias voltages.

With the side rods on "ground" potential (i.e., the potential of the plasma chamber) the source running on $^{13}CH_4$-gas was tuned for highest currents of $^{13}C^{5+}$ ions. The charge state distribution, measured at 5 kV extraction voltage and RF power of 230 W showed carbon ions of all charge states, where usage of the biased disk was crucial.

It turns out that the source tuning in this particular set-up is comparable to other ECRISs, of course with substantially lower currents. With the biased disk set at some negative voltages the $^{13}C^{5+}$ ion currents were measured to see the effect of the rods.

The first test was to have all side electrodes on a voltage of $U_{rod} = +10$ V, 0 V, and $-5$ V, respectively. The highest $C^{5+}$ ion currents for the three cases were 540, 360, and 320 nA $C^{5+}$, respectively. It proved that all side electrodes should be biased on a positive voltage, confirming the trend found in the earlier measurements.[8]

TABLE I. Measured ion currents for several values of bias voltages on various sets of rods inside Bio-Nano ECRIS.

| "Injection" rods (V) | | "Extraction" rods (V) | | Biased disk (V) | $C^{5+}$ (nA) |
|---|---|---|---|---|---|
| 0 | | 18 | | −29 | 720 |
| 17 | | 0 | | −29 | 730 |
| 17 | | 0 | | −36 | 780 |
| 14 | | 7 | | −21 | 770 |
| 0 | | 0 | | −29 | 530 |
| 0 | | 0 | | −35 | 740 |
| One | Others | | | | |
| 30 | 0 | 0 | | −29 | 770 |
| 0 | 25 | 0 | | −29 | 780 |
| 0 | 0 | 0 | | −29 | 530 |
| | | One | Others | | |
| 0 | | 28 | 0 | −29 | 720 |
| 0 | | 0 | 0 | −29 | 540 |

In a preliminary set of measurements it was found that biasing rods at locations between the six rods shown in Figure 1 had little or no influence on the source operation. A similar finding at the Kei source was reported[8] earlier.

### B. Biasing a group of three rods

Since each rod has a separate electrical feed-through, one can apply voltages to different combinations of rods. The rods were split into two groups; one called "extraction rods" corresponding to the triangle visible at the extraction end plate, see Figure 1, and the other called "injection rods."

From the values of measured $^{13}C^{5+}$ ion currents in Table I one can conclude that the effect of biasing either the group of injection rods at 17 V or the group of extraction rods at 18 V has approximately the same effect: the output increases by 35% to 40%.

### C. Biasing one rod only

The most interesting observation is the result of biasing only one rod of a group of three rods. The effect is very much comparable to the case of biasing two rods of a group or even all the three rods of a group, as can be seen from the table. The (significantly) best result was obtained with one or two rods from the injection group; the output of $^{13}C^{5+}$ ions increased by about 45%. On first sight all these observations are against first order expectations in terms of linear behaviour.

### D. Biased rod adding to biased disk

Unfortunately, during these measurements no good answer could be given to this important question. In fact only one measurements with varying various source parameters was done showing that with $U_{BD} = -35$ V while $U_{rods} = 0$ V about the same output of $C^{5+}$ ion current was obtained (see Table I) as for a few combinations with $U_{inj\,rods}$ and/or $U_{extr\,rods}$ positively biased.

## E. Explanation and discussion

A simple explanation in terms of reduction of the radial leaks is not easy. On first glance one would expect to see a lower effect of biasing three rods as compared to biasing six rods. However, by taking some symmetry reasons into account three rods possibly could be as effective as six rods are. But such a symmetry argument cannot explain the fact that biasing just one rod has approximately the same effect as biasing a group of three rods.

One should keep in mind that a change of any external parameter will bring about that the properties of the ECRIS plasma will change as well. In fact, due to a reduction of the ion leakage, the plasma potential will change because it is a selfconsistent parameter, regulating that the total of the charge (of ions and electrons) leaking in all possible directions out of the plasma equals zero.

The conclusion is that for the given set-up the effect of (one or several) biased rods is definitely apparent.

## III. EXPERIMENTS IN THE ATOMKI ECRIS

Based on the results described above with the Bio-Nano ECRIS it was decided to perform some measurements with the Atomki ECRIS at Debrecen. A simple modification for the installation of one (movable) rod could easily be achieved; installation of three (or even six) rods, each of them separately insulated, would definitely be impossible taking into account the small diameter (58 mm) of the plasma chamber as well as the heavy source usage.

The Atomki ECR ion source has a flexible configuration.[11] For this experiment settings for the production of highly charged ions were chosen. The intention is to check the output as a function of rod position and bias. Here are some details:

- The RF system operates at 14 GHz always at a power setting of 800 W.
- The coil currents are set at maximum; the ECR surface in the midplane is at r = 15 mm.
- The plasma chamber with internal diameter of 58 mm is installed.

The experiments have been performed with two different side electrodes, described below.

### A. Movable tantalum rod with variable length at pole position

The rod with diameter 5 mm is attached with an insulator to the injection plate; it can be moved over range of z = 0 to 146 mm, where z = 0 mm corresponds to the injection plate (see Figure 2). The distance from rod to the wall is about 8 mm. Note that the midplane (magnetic field of the coils) is located at z = 135 mm.

It was decided to operate this source with $^{22}$Ne gas without any gas mixing. All charge states can be detected unambiguously. In the analysis the highly charged $^{22}$Ne$^{8+}$ ion beam currents were used. In the Atomki ECRIS the use of the biased disk is extremely important. The best spectrum

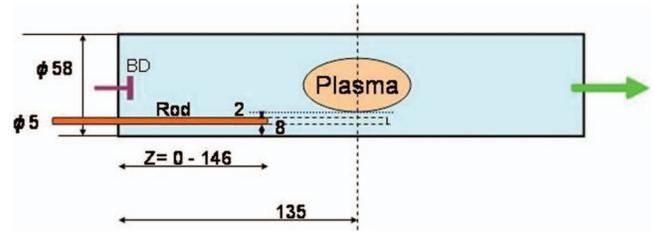

FIG. 2. Set-up of the Atomki ECRIS with one single movable rod at the side. All sizes are in mm, the drawing is not to scale.

($U_{BD} \approx -150$ V) showed a $^{22}$Ne$^{8+}$ ion current of 23 $\mu$A, and with $U_{BD} = 0$ V the best $^{22}$Ne$^{8+}$ ion current is about 4 $\mu$A. One "feature" apparent as soon as the rod was moving closer to the plasma is outgassing – likely due to heating of the probe – and thereby giving a tendency for instabilities. Measurements were performed taking ample time for getting a stable situation, with most parameters fixed. In the following some of the results are summarized.

(1) A scan was made (while $U_{BD} = 0$ V) with the rod at positions between z = 0 and 88 mm and at each position $U_{rod} = -10$, 0, and +10 V. The highest $^{22}$Ne$^{8+}$ current (=5.0 $\mu$A) was measured for the rod at position z = 84 mm and $U_{rod} = +10$ V; at the same rod position and $U_{rod} = 0$ V a lower $^{22}$Ne$^{8+}$ current (=4.6 $\mu$A) was found. This observation corresponds in a way to the earlier experiment with BioNano ECRIS. In all these and following cases the rod *current* was positive for $U_{rod} > +5$ V. That means a transport of ions from rod to plasma and/or a transport of electrons from plasma to rod which is in agreement with the expected effect.

(2) A scan was made (while the disk was at optimized position of 15 mm and the rod at fixed position of z = 87 mm) *with $U_{BD}$ varying between 0 and –250 V and $U_{rod}$ between 0 and +25 V*. The highest $^{22}$Ne$^{8+}$ current (=14.7 $\mu$A) was found for $U_{BD} = -250$ V and $U_{rod} = +20$ V; this combination seems here to form a local maximum, see Table II. This observation is in the trend of the earlier experiment with BioNano ECRIS. At the best combination a CSD was measured, see Figure 3.

(3) A scan was made (*while the BD was at optimized position and $U_{BD} = -154$ V*) with the rod at position from

TABLE II. For several combinations of the biased disk (BD) voltage and rod voltage the ion currents (in nA) are shown.

| | | Rod @ 87 mm, BD @ 15 mm | | | | |
|---|---|---|---|---|---|---|
| | $U_{Rod}$ (V) | 2.6 | 7 | 10 | 20 | 25 |
| $U_{BD}$ (V) | $i$ ($^{22}$Ne$^{8+}$) | | | | | |
| −250 | | 12.0 | 11.9 | 12.5 | 14.7 | 14.2 |
| −220 | | 12.1 | 12.1 | 12.7 | 14.5 | 14.2 |
| −190 | | 12.2 | 12.1 | 12.8 | 14.4 | 13.9 |
| −160 | | 12.2 | 12.1 | 13.0 | 14.3 | 13.6 |
| −130 | | 12.1 | 12.0 | 12.9 | 14.2 | 13.6 |
| −100 | | 12.1 | 11.8 | 12.7 | 14.3 | 13.9 |
| −70 | | 12.2 | 11.5 | 12.6 | 14.2 | 14.2 |
| −40 | | 10.9 | 11.3 | 11.9 | 13.9 | 13.9 |
| −10 | | 5.7 | 5.5 | 5.9 | 7.9 | 9.2 |

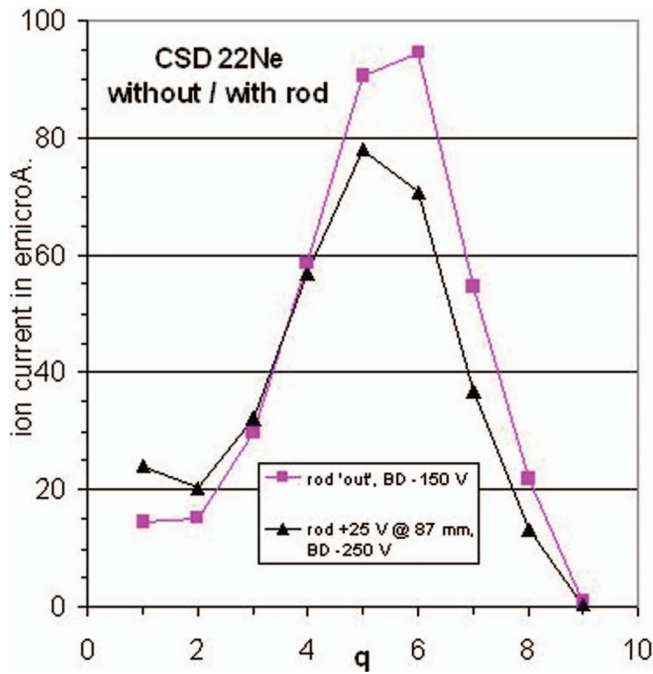

FIG. 3. CSD for the best settings in Table II. For comparison the CSD is taken with rod "off," i.e., rod at z = 0 position.

z = 0 to 88 mm with $U_{rod} = 0$ V. Here the highest $^{22}Ne^{8+}$ current (=23 μA) was for the rod at position 0; the current was approximately linearly decreasing with rod po sition. This observation corresponds not at all to the earlier experiment with BioNano ECRIS.

Of course, the result of scan #3 seems on first sight disappointing. But since the results of the other scans are a bit positive, the results likely are the summation of a positive effect of the biased rod (see Table II) and the negative effect of the location of the rod, so close (2 mm) to the ECR zone as becomes apparent from Figure 2.

### B. Movable side electrode with fixed length

As a natural extension of the first experiment and also to geometrically separate the effect of the disk and the rod a rapid second measurement was performed with a modified set-up (Figure 4). A short movable rod (50 mm of length) as close as possible to the wall was located at one of the

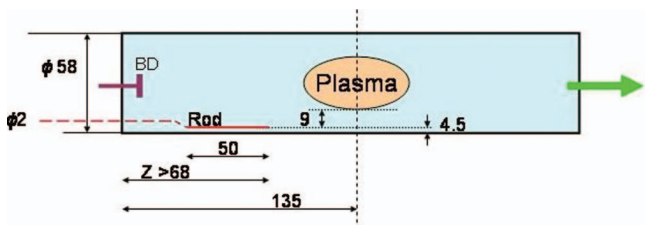

FIG. 4. The Atomki ECRIS with one short movable rod as close as possible to the side wall (schematic; sizes are in mm).

After extensive outgassing using again a high-charge state setting at 800 W RF power, the rod was moved closer to the plasma but it turned out to be extremely difficult to stabilize the plasma for rod positions larger than z = 80 mm. At this position a few CSDs were measured as well as a map where $^{22}Ne^{8+}$ currents were recorded as a function of Biased Disk voltage and rod voltage.

From these measurements it is clear that for the given ECRIS the BD is playing a dominant role, even with the rod installed. Note that the experiment had to be ended because the rod showed a short to the wall. After opening it appeared that the rod end was melted.

## IV. DISCUSSION

The influence of application of biased side electrodes was checked at three sources. It was clearly proven that a pos itively biased side electrode decreases the ionic losses, i.e., one obtains higher HCI currents with an (optimised) positive voltage than at zero voltage.

In the Kei-source the apparent effect is to move the charge state distribution towards higher charge states, even while using optimized biased disk settings. In the Bio-Nano ECRIS the effect is significant, although a biased disk can almost reach the same improvement of highly charged ions output depending on the (high vacuum) situation. In the Atomki
ECRIS the effect also can be demonstrated in case the BD is switched off. However, with BD in operation, the improvement due to biasing one side rod can be demonstrated but it is balancing negatively with that of the rod inside the plasmachamber, likely due to the inherent heating of the side rod by the applied high power.

This brings about the following outlook for application of one (or more) biased side electrode.

It likely will be advantageous if there is no possibility for applying an axial biased disk, e.g., if an oven or other large-size tool must be installed on axis. A water-cooled variant of a movable side-rod deserves a test in any high-power ECRIS. Also, it still could be applied in case it is integrated into the chamber wall and well cooled.

### ACKNOWLEDGMENTS

We are indebted to Y. Kato, L. Kenez, H. Minezaki, M. Muramatsu, K. Oshima, and Y. Yoshida for advice and help during various stages of the experiment. The participation of two of the authors (R.R. and S.B.) in the present work was partly supported by the TAMOP 4.2.2.A-11/1/KONV-2012-0036 project, which is co-financed by the European Union and European Social Fund.